\documentstyle[prl,twocolumn,aps]{revtex} 
\newcommand{\tr}{{\rm tr}}
\begin{document}
\draft
\title{Traces and determinants of strongly stochastic operators}
\author{C. P. Dettmann}
\address{Niels Bohr Institute, Blegdamsvej 17, 2100 K{\o}benhavn \O, Denmark
\\dettmann@nbi.dk}
\date{\today}
\maketitle
\begin{abstract}
Periodic orbit theory allows calculations of long time properties of
chaotic systems from traces, dynamical zeta functions and spectral
determinants of deterministic evolution operators, which are in turn
evaluated in terms of periodic orbits.
For the case of stochastic dynamics a direct numerical evaluation
of the trace of an evolution operator is possible as a multidimensional
integral.  Techniques for evaluating such path integrals are discussed.
Using as an example the logistic map $f(x)=\lambda x(1-x)$ with moderate
to strong additive Gaussian noise, rapid convergence is demonstrated for
all values of $\lambda$ with strong noise as well as at fixed $\lambda=5$
for all noise levels.
\end{abstract}
\section{Introduction}
Periodic orbit theory is a remarkable tool in classical~\cite{AAC,Cv,Ga} and
quantum~\cite{Gu,GA} chaotic systems, permitting the evaluation of long
time properties, such as escape rates, dynamical averages and energy levels
in terms of short unstable recurrent motions, that is, periodic orbits
or ``cycles''.  In classical hyperbolic systems with known topology,
such as the repeller
of the map $5x(1-x)$ discussed below, convergence can be impressive: the escape
rate is computed to 9 digits from the 8 cycles of period 4 or less.
The lowest energy levels of helium~\cite{WRT} are computed to an accuracy
far better than would be expected from a semiclassical approximation.
Even when there is intermittency, hence hyperbolicity is lost, it is possible
to get sensible results using special techniques in both the
classical~\cite{Da,DD} and quantum~\cite{Ha} cases.

Recently the theory has been extended to classical systems with weak
additive noise, using Feynmann diagrams~\cite{CDMV1} or smooth
conjugacy techniques~\cite{CDMV2}.  There are a number of motivations
for such extensions: noise at some level is present in all physical systems;
it regularizes the theory, replacing Dirac $\delta$-functions by smooth kernels
(see below) and fractal distributions by smooth functions; there is also
some hope that the noise may effectively truncate the theory, rendering
irrelevant contributions from periodic orbits longer than the finite memory
of the system.

The result of these investigations is a weak noise perturbation theory,
representing the trace of the evolution operator and derived quantities
as a power series expansion in $\sigma$, the noise level.  The coefficients
are combinations of higher derivatives of the map evaluated at the periodic
orbits of the deterministic unperturbed system.  Numerically, the
coefficients themselves converge at a similar rate to the classical
periodic orbit theory, but the power series in $\sigma$ is useful only
for weak noise, say $\sigma<0.03$, suggesting the following question,
the subject of this paper:

{\it To what extent does periodic orbit
theory survive strong noise, and how fast does it converge?}

Strong noise differs qualitatively from weak noise in a number of respects:
The stochastic dynamics is equally close to many slightly different
deterministic dynamical systems, so the concept of a unique
perturbation theory becomes less defined, in addition to the lack of
convergence of such a theory.  Also, Gaussian noise has no preferred status;
for weakly stochastic systems, all types of noise distributions with a
given variance $\sigma^2$ are identical to order $\sigma^2$.

The approach taken here is that the relevant quantity, the trace of an
evolution operator, is evaluated numerically, using very little detailed
information about the dynamics, in particular without reference to periodic
orbits.  The method is general enough to include any type
of dynamics (hyperbolic, intermittent, attracting) and uncorrelated
noise, subject to smoothness of both dynamics and the noise distribution,
with the latter decaying exponentially at large distances.  Here, as in
Refs.~\cite{CDMV1,CDMV2} the noise is additive, but this is not a
necessary condition.

From the trace, it is straightforward to construct the spectral determinant,
and hence highly convergent expansions for escape rates and dynamical
averages, in the spirit of cumulant expansions, as in standard periodic
orbit theory.  This has some similarities to Ref.~\cite{Wi}, where various
approximations to the quantum trace are compared.

Section~\ref{form} outlines the formalism required for the calculation,
in particular casting the trace as a multidimensional integral.
Section~\ref{sint} discusses numerical approaches for evaluating this
integral.  Finally, the results and their ramifications are discussed in
Section~\ref{res}.

\section{Formalism}\label{form}
The goal is to determine the long time properties of stochastic dynamical
systems, here one dimensional maps with additive noise:
\begin{equation}
x_{n+1}=f(x_n)+\sigma\xi_n\;\;,\label{Lang}
\end{equation}
where $f(x)$ is a known function, for example the logistic map
\begin{equation}
f(x)=\lambda x(1-x)\;\;,\label{map}
\end{equation}
$\sigma$ is a measure of the strength of the noise, and
$\xi_n$ are independent identically distributed random variables with
unit variance,
\begin{equation}
\langle \xi_m\xi_n \rangle=\delta_{mn}\;\;,
\end{equation}
such as a normalized Gaussian distribution.  The methods used
here are equally applicable to $\sigma$ and $\xi$ that depend on $x$, and
non-Gaussian noise distributions.

Instead of the Langevin form (\ref{Lang}) it is more convenient to consider
the discrete Fokker-Planck equation for a probability distribution
$\rho(x)$ transported by the dynamics and diffusing due to the noise:
\begin{equation}
\rho_{n+1}(x)\equiv{\cal L}[\rho_n](x)
=\int \delta_\sigma(x-f(x'))\rho_n(x')dx'
\end{equation}
where $\delta_\sigma(y)$ is the noise kernel, for example
\begin{equation}
\delta_\sigma(y)=\frac{e^{-y^2/(2\sigma^2)}}{\sigma\sqrt{2\pi}}\;\;,
\end{equation}
reducing to a Dirac $\delta$ in the deterministic $\sigma=0$ limit.

Long time properties of the dynamics are obtained from the leading
eigenvalue(s) of the linear evolution operator ${\cal L}$, which are
(the inverses of) solutions of the characteristic equation
\begin{equation}
\det(1-z{\cal L})=0\;\;.\label{det}
\end{equation}
For example, the probability of a point initially in an open system
remaining there after $n$ iterations is typically proportional to
$e^{-\gamma n}$ where the escape rate $\gamma$ is related to the
leading zero $z_0$ by
\begin{equation}
\gamma=-\ln z_0\;\;.
\end{equation}
Dynamical averages and diffusion coefficients can be obtained from the
leading zero of appropriately weighted evolution operators~\cite{AAC,Cv,Ga}.

The spectral determinant~(\ref{det}) of an infinite dimensional operator
may be defined by its cumulant expansion in powers of $z$, using the
matrix relation $\ln\det=\tr\ln$ and Taylor expanding the logarithm:
\begin{eqnarray}
\det(1-z{\cal L})&=&
\exp\left(-\sum_{n=1}^{\infty}\frac{z^n}{n}\tr{\cal L}^n\right)\nonumber\\
&=&1-z\tr{\cal L}+\frac{z^2}{2}\left[(\tr{\cal L})^2-\tr{\cal L}^2\right]
-\ldots\nonumber\\
&\equiv&\sum_{n=0}^{\infty}C_nz^n\label{cum}
\end{eqnarray}
$C_n$ may be obtained from all the traces $\tr{\cal L}^m$
with $m\leq n$, and an approximation for $z_0$ is obtained by numerical
root finding on the $n$'th degree polynomial given by the truncation of
the determinant.  The above expression for $C_n$ quickly gets complicated;
it is easier to expand the derivative
\begin{equation}
-z\frac{d}{dz}\det(1-z{\cal L})
=\det(1-z{\cal L})\sum_{n=1}^{\infty}z^n\tr{\cal L}^n\;\;,
\end{equation}
which leads to the recursive equation
\begin{equation}
C_n=\frac{1}{n}\left(\tr{\cal L}^n-
\sum_{m=1}^{n-1}C_m\tr{\cal L}^{n-m}\right)\;\;.
\label{rec}
\end{equation}

The trace is straightforward to write down as an $n$-dimensional integral,
a discrete periodic chain reminiscent of a path integral, obtained in
Ref.~\cite{SWM},
\begin{equation}
\tr{\cal L}^n=\int\prod_{j=0}^{n-1}dx_j
\prod_{j=0}^{n-1}\delta_\sigma(x_{j+1}-f(x_j))\label{int}
\end{equation}
where the index $j$ is cyclic, so $x_n=x_0$.  In the noiseless ($\sigma=0$)
limit, the integrand is a product of Dirac $\delta$-functions, and the
trace is given by a sum over the fixed points of $f^n$, that is, the
$n$-cycles of $f$.  In Refs.~\cite{CDMV1,CDMV2} the weak noise limit is
obtained by a saddlepoint expansion of the integral around these cycles.
Here, the integral is performed numerically, up to $n=5$, as described in the
following section.

\section{Numerical methods}\label{sint}
The required quantities, $\tr{\cal L}^n$, are $n$-dimensional integrals,
which in the case of weak noise ($\sigma\ll 1$) have a large number of
sharp peaks surrounding the periodic points of the deterministic map.
Obtaining an accurate numerical estimate of the integral for any $n>2$
seems prohibitively difficult, since Monte Carlo approaches take too
long to converge, and direct integration schemes require a small step
size, but cover a large configuration space.  See Ref.~\cite{PTVF} for more
discussion.

In the case of Gaussian (or similar) noise and smooth dynamics $f(x)$ 
the integrand is smooth and decays exponentially
fast at the boundaries.  This in turn implies that the simplest possible
integration algorithm, summing the integrand at a cubic array of coordinate
values, converges faster than any power of the step size, and is typically
exponential once the step size is smaller than $\sigma$.

This remarkable convergence rate for smooth, exponentially decaying integrands
follows from the observation that by multiplying the terms near the boundary
by appropriate factors, it is possible to obtain algorithms of higher and
higher order in the step size~\cite{PTVF}.  Exponentially decaying
integrands are impervious to any such coefficients, and so converge
faster than any power of the step size.

Note also, that having chosen, say $x_0$ and $x_1$, and the argument of
the exponential, $-(x_1-f(x_0))^2/(2\sigma^2)$ happens to be too small,
it is not necessary to consider the other $x_j$.
This provides a very substantial saving in time for $n>2$.

Finally, the integral is symmetric under a cyclic interchange of the
$x_j$: this implies an additional saving of a factor $n$.  The logic
required here is not trivial since the contribution differs depending
on whether some of the $x_j$ are identical.  For example, for $n=4$,
choose two values $x_{min}$ and $x_{max}$ beyond which there is no
possible contribution.  Then sum $x_{min}\leq x_0\leq x_{max}$, defining
$x_0$ to be the largest of the $x_j$, and the one occurring first, if more
than one are maximum.  Sum $x_{min}\leq x_1\leq x_0$, checking that the
argument of the exponential is not too small.  Sum
$x_{min}\leq x_2\leq x_0$, again checking the argument of the exponential.
Then sum $x_{min}\leq x_3<x_0$, and multiply each contribution by 4.
If $x_2=x_0$ the $x_j$ could form a 2-cycle repeated twice, so when
$x_3=x_1$ count the term twice instead of four times, and stop the sum
over $x_3$ to avoid double counting.  Finally, the repeated fixed point
$x_0=x_1=x_2=x_3$ has been excluded, so sum this explicitly and
count it once.  The case $n=5$ is simpler as there is only a repeated
fixed point, but there are more possibilities for which of the $x_j$
are maximum.

Even with the above short cuts, large $n$, small $\sigma$, and stringent
precision requirements can lead to sums of $10^9$ terms.  This means it is
advisable to group them in size (using the argument of the exponential)
as they are summed, then combine the groups from smallest to largest
to minimize roundoff error.

Given the above algorithm the step size $h$ is decreased until two
successive estimates agree to within a specified precision (for example
10 digits).  Since the amount of time increases as $h^{-n}$ the optimal
sequence is probably $h_j=h_0 e^{-j/n}$.  Note that large initial
values of $h$ can lead to a zero result as the entire contribution
region may be missed. 

With the above algorithm, calculation of the trace up to $n=5$ with
$\sigma\geq0.01$, and $n=6$ for somewhat higher values of $\sigma$,
is feasible for the case of Gaussian noise and smooth one dimensional
dynamics.

\section{Results}\label{res}
The logistic map $f(x)=\lambda x(1-x)$ for various values of $\lambda$
exhibits most of the behaviors observed in one dimensional maps.  For all
$\lambda\geq 1$ any initial $x$ outside the range $[0,1]$ ends up at $-\infty$,
while the behavior of points within this range depend of $\lambda$ as
follows: For $0\leq\lambda\leq1$, the point $x=0$ is a stable fixed point,
marginally so at $\lambda=1$, and then unstable for $\lambda>1$.  For
$1\leq\lambda\leq3$, the fixed point $x=1-1/\lambda$ is stable, and then
bifurcates to a stable cycle of period 2.  This cycle in turn becomes
unstable, bifurcating to a 4-cycle, then an 8-cycle, and so on, to
$\lambda\approx 3.57$ at which point a chaotic attractor forms.  The period
doubling cascade in the presence of weak noise may be described by the
renormalization approach of Ref.~\cite{SWM}.  At larger
values of $\lambda$ more stable cycles are created, including a 3-cycle
which is stable at $\lambda=3.84$, leading to a pattern of alternating
stable ``windows'' surrounded by non-attracting unstable cycles and chaotic
attractors containing many unstable cycles.  At $\lambda=4$ the attractor
fills the interval $[0,1]$, and in this case, the Ulam map, the
dynamics is exactly solvable.  For $\lambda>4$ almost all initial conditions
leave the interval, but infinitely many unstable cycles remain, forming
a fractal repeller, with a well defined escape rate.

Imposing additive noise to the logistic map leads to escape for all
$\lambda>0$, although this may be very unlikely if $\sigma$ is small.
At $\lambda=2$, for example, every point (except the endpoints) is
attracted to the stable fixed point at $x=1/2$, and the noise must
move the trajectory out of the interval to escape.  In cases like this,
the stochastic behavior is analogous to quantum tunneling, and is
exponentially suppressed for small $\sigma$.  At bifurcation points,
including $\lambda=1$, the stability of the relevant
cycles is marginal, leading to intermittency.  Marginal cycles are
difficult to treat using cycle expansions, and it is
one of the goals of this work to understand how this poor convergence
is modified by the presence of noise.

The results of numerical evaluation of $\tr{\cal L}^n$ up to $n=5$ are shown
in Tab.~\ref{conv}.  The spectral determinant is evaluated using
(\ref{rec}), and $C_5$, the coefficient of $z^5$ is noted.  Since for
the parameters shown the first zero of the determinant is close to $1$,
$-\log_{10}|C_5|$ gives roughly the number of significant digits of $z$,
and hence the escape rate, evaluated to $n=4$.  It also gives the
approximate range of $z$ over which the $n=4$ approximation is valid.

It is seen that, for the trivial case $\lambda=0$, corresponding to pure
noise, and for strong noise $\sigma=1$, the calculation is limited by the
double precision arithmetic: evaluation of the trace beyond $n=4$ is
superfluous at this level of precision.  Almost as precise is the case
$\lambda=5$ which has a repeller with complete binary symbolic dynamics
in the absence of noise, and hence is an ideal candidate for cycle
expansion methods.  Nine significant digits are obtained at $n=4$,
corresponding to just 8 cycles.  The presence of noise makes methods
based on enumerating these cycles more difficult~\cite{CDMV1,CDMV2},
but convergence is rapid at any noise level.

The other cases, where escape is induced by the presence of noise, do
rather poorly for small noise.  The significance of
$\lambda=1,2,3,3.57,3.84,4$ are discussed above; the other values in
Tab.~\ref{conv} are $\lambda=3.5$ which contains a stable 4-cycle,
and $\lambda=3.72$ which is not near any large stable window, and
numerically exhibits a chaotic attractor, although mathematical proof
is difficult.  The nature of the underlying attractor seems to have
little effect on the rate of convergence, except that the intermittent
case ($\lambda=3$ and particularly $\lambda=1$) is divergent at $\sigma=0.01$
to this level of approximation; the escape rate probably converges at
impossibly large $n$, either for the current numerical approach, or for
standard cycle expansion techniques.  In the other cases, particularly
towards larger $\lambda$ the expansion appears to be converging, albeit
slowly.

In conclusion: What are the optimal methods for determining
the long time properties of stochastic systems? The strong noise case
is best treated by numerical evaluation of the trace, described here, 
requiring little knowledge of the underlying dynamics.  The elements
of periodic orbit theory, traces and determinants indeed survive
strong noise, and converge rapidly, without reference to periodic orbits.
The weak noise case depends on this underlying dynamics: for the hyperbolic
case ($\lambda>4$), the cycle perturbation theory of~\cite{CDMV1,CDMV2} or
numerical evaluation;
for noise induced escape from a strongly chaotic attractor
($\lambda\approx4$), the analytic methods of~\cite{Re}; and for
tunneling from
a stable fixed point, analytic approaches analogous to quantum
mechanics.  The intermittent case with weak noise remains an open problem;
the results here show that weak noise does not substantially
regularize cycle expansions of intermittent systems, at least with respect
to the rate of convergence. 

The author is grateful for helpful discussions with V. Baladi and
P. Cvitanovi\'{c}.

\begin{table}
\begin{tabular}{|l|c|rrrrr|}
&&&&$\sigma$&&\\
$\lambda$&Type&0.01&0.03&0.1&0.3&1\\\hline
0&Pure noise&12.7&12.7&12.4&12.7&12.6\\
1&Intermittent&-2.3&-0.8&1.2&3.8&8.5\\
2&Stable 1-cycle&2.5&2.2&2.1&5.9&11.8\\
3&Bifurcation&-0.3&0.7&2.8&7.4&13.2\\
3.5&Stable 4-cycle&0.3&1.4&3.4&7.8&13.2\\
3.57&$\infty$-cycle&0.4&1.1&3.6&7.8&13.3\\
3.72&Chaos&1.5&1.4&4.1&8.0&13.4\\
3.84&Stable 3-cycle&1.6&2.4&4.6&8.1&13.4\\
4&Ulam map&2.2&2.9&4.9&8.2&13.8\\
5&Repeller&9.2&9.1&8.4&9.1&13.3
\end{tabular}
\caption{Convergence of the spectral determinant, as measured by
$-\log_{10}|C_5|$, where $C_5$ is the coefficient of $z^5$ in the cumulant
expansion (\protect\ref{cum}) for various types of dynamics of the logistic
map (\protect\ref{map}).
Larger numbers imply faster convergence, giving roughly the number
of converged digits in the escape rate calculated to $n=4$.
\label{conv}}
\end{table}

\end{document}